\documentclass[prb,twocolumn,,floatfix]{revtex4}
\usepackage{graphicx}
\usepackage{bm}
\usepackage{amssymb}
\usepackage{color}

\newcommand{\be}{\begin{equation}}
\newcommand{\ee}{\end{equation}}

\def \be{\begin{equation}}
\def \ee{\end{equation}}
\def \ba{\begin{array}}
\def \ea{\end{array}}
\def \bea{\begin{eqnarray}}
\def \eea{\end{eqnarray}}

\begin{document}

\title{Stability of nodal quasi-particles in superconductors with coexisting
orders}
\author{E. Berg, C-C. Chen, and S. A. Kivelson}
\affiliation{Department of Physics, Stanford University, Stanford CA 94305-4045, USA}

\begin{abstract}
We establish a condition for the perturbative stability of zero energy nodal
points in the quasi-particle spectrum of superconductors in the presence of
coexisting \textit{commensurate} orders. The nodes are found to be stable if
the Hamiltonian is invariant under time reversal followed by a lattice
translation. The principle is demonstrated with a few examples.
%Experimental
Some experimental implications of various types of assumed order are
discussed in the context of the cuprate superconductors.
\end{abstract}

\date{\today}
\maketitle

%\emph{Introduction - }
One of the most distinct properties of unconventional (\textit{e.g.} d-wave)
superconductors is the possible existence of nodal quasi-particles (QP) with
a linear Dirac-like spectrum in two dimensions, or lines of nodes in three
dimensions. As the QP's dominate the low energy asymptotics of many physical
properties, they are an essential feature of the state. A nodal
%quasi-particle
QP spectrum in two dimensions has a linearly vanishing density of states %that
%vanishes linearly
at zero energy which %leads to
produces several clear experimental
fingerprints, \textit{e.g.} a universal contribution to the thermal
conductivity\cite{fradkin,DurstLee}, a linear temperature dependence of the
penetration depth at low temperatures\cite{lin_T_theory}, and a $\sqrt{H}$
dependence of the heat capacity in a magnetic field\cite{volovik}. %The Dirac
The spectrum can also be observed %using spectroscopic probes, such as
directly in ARPES and
STM.

The cuprate high temperature superconductors are known to be d-wave
superconductors, and the properties of their nodal QPs have been carefully
explored in many experiments. Many theories have been proposed to account
for the mechanism of superconductivity %in the cuprates
and the nature of the anomalous normal state. Some of these theories involve
another type of ordering, that can either compete with superconductivity,
coexist with it, or enhance it. Some examples, which we will treat
explicitly, are antiferromagnetism, charge or spin density wave orders (e.g.
\textquotedblleft stripes\textquotedblright ), and at least two forms of
time-reversal symmetry breaking orbital antiferromagnetism:
\textquotedblleft d-density wave" (dDW) with an ordering vector $\mathbf{q}%
=(\pi ,\pi )$ %pattern
\cite{DDW}, and ``Varma loops''\cite{Varma} which does not break the
translational symmetry of the crystal, \textit{i.e.} $\mathbf{q}=(0,0) $,
but at the same time has no net orbital moment. Various experimental and
theoretical studies have %, in recent years,
provided evidence (sometimes conclusive, sometimes suggestive) of the
existence of such \textquotedblleft competing ordered states"\cite%
{TRbreaking}.%
%a different pattern\cite{Varma}.
%Many of these
%orders appear naturally in mean-field solutions of microscopic
%models for electrons with repulsive interactions, such as the
%Hubbard and t-J models, and some of them emerge naturally from
%numerical solutions of these models\cite{WhiteScalapinoStripes}.
%On the experimental side, static spin and charge unidirectional
%(or ``stripe") order has been observed in some cuprate
%superconductors, namely LaBaCuO\cite{LaBaCuO} and
%LaNdCuO\cite{LaNdCuO}, close to a doping level of $x=1/8$. In
%these materials, the stripe order seems to compete with
%superconductivity, in the sense that the superconducting $T_c$ is
%suppressed very strongly at the doping where the stripe order is
%strongest. However, recent transport
%experiments\cite{TranquadaLBCO} have shown evidence for strong
%2d-like superconducting fluctuations that seem to coexist with the
%stripe order. In recent Kerr effect measurements\cite{AharonKerr}
%in YBCO-123, a finite signal was detected, with an onset
%temperature close to the pseudogap temperature. The signal extends
%well into the superconducting phase. These experiments, which are
%consistent with earlier $\mu$SR\cite{mSR} and polarized neutron
%scattering\cite{Bourges} measurements, imply that a time reversal
%symmetry breaking order exists in YBCO, and that this novel type
%of order coexists with the superconducting order at least in some
%region of the phase diagram.

These %striking
observations lead us to address the following question: how is the
quasi-particle spectrum of a d-wave superconductor generically affected by
the presence of a coexisting order?
%Specifically, since the nodes typically
%dominate the low-energy electronic properties, it is important to know
%whether they are stable with respect to the addition of such order.
%Naively,
Since the nodal QPs are gapless,
% lie at zero energy, they are not protected by an energy gap,
%so any
%there likely
it is not surprising that there exist (as we shall show) certain classes of
infinitesimal perturbations that can change them qualitatively, either by
gapping them, or by expanding the gapless locus in k-space from a point to a
closed line (a ``Fermi surface pocket'').

The question of the stability of the nodes was addressed in several previous
studies, in particular in the presence of a general spin-orbit coupling\cite%
{sato}, in d-wave superconductor with several specific types of coexisting
order\cite{checkerboard,dwavechecker,vojta,tesanovic,Granath,Subir1,Subir2} and in a %magnetic
vortex lattice\cite{Halperin,ashvin,oskar}.
%In the latter case, because time reversal symmetry is
%broken, the QP spectrum is %generically
%gapped unless very special (non-generic) symmetries are % is present
%assumed.%
% protected if the spectrum is linearized\cite%
%{Halperin,ashvin}, or, more generally, in the presence of
%particle-hole and parity symmetries\cite{oskar}.

In this paper, the question of the stability of the nodal QPs in the
presence of a competing order of weak to moderate strength is addressed in
the mean-field approximation. Specifically, we consider the quasi-particle
spectrum of a 2d partially filled band in the presence of a uniform pairing
field with d-wave symmetry and a second effective field, representing the
competing order, which couples to a fermion bilinear and which we will
consider as a ``perturbation.''
%We consider a mean-field 2d superconductor on a lattice with Dirac nodes,
%and some coexisting, two-fermion order parameter that can break any symmetry
The competing order can break any symmetry, such as translation symmetry or
time-reversal symmetry; the only condition %on the coexisting order
is that it is \textit{commensurate}, \textit{i.e.} the period of the
%new order
ordered state is a rational multiple of the original lattice constant. Under
these assumptions, we show that if the perturbation does not ``nest" %two
any pair of nodes (\textit{i.e.}, no two nodal k-points are coupled directly
by the perturbation), and if the perturbation is invariant under time
reversal or time reversal followed by a %lattice
translation, then the nodes are stable %over a
%finite range of magnitude  perturbation.
at least until the strength of the perturbation exceeds a non-zero critical
value. However, if the perturbation is not invariant under any such relative
of time reversal symmetry, %followed by any
%translation,
then the nodes can become gapped or can be shifted from zero energy for any
infinitesimal amount of perturbation. (Even where the nodes are
perturbatively stable, as the perturbation increases in strength, the
location of the nodes in k-space generally shifts. In many cases, this
eventually leads to a situation in which a pair of nodes meet or satisfy a
more general nesting condition; %, in which case
%cases, This means that as long as the
%perturbation is not strong enough to move the nodes until they become
%nested, the nodes cannot be gapped and are pinned to zero energy. When two
%nodes ``meet" in the down-folded Brillouin zone (which is equivalent to
%nesting),
then, they can annihilate each other %and become gapped.
leading to a gapped spectrum.) % If the
%perturbation is not invariant under time reversal followed by any
%translation, then the nodes can become gapped or be shifted from zero energy
%for any infinitesimal amount of perturbation.

%This paper is organized as follows. First, we describe the argument for the
%stability of the nodes under the conditions described above. We then bring
%some examples of coexisting orders that satisfy the conditions, and others
%that do not. Solving the Bugoliubov-de Gennes equations numerically, we
%demonstrate that in the cases that these conditions are not met, the nodes
%are generically unstable. The experimental implications of these results are
%then discussed.

We now turn to the derivation of our main result.
%, namely the stability of the nodal
%points.
Consider a two dimensional d-wave superconductor on a lattice, whose
spectrum we assume is well approximated by a uniform BCS mean-field
Hamiltonian, $\mathcal{H}_0$. Suppose that we add a periodic perturbation
%which is assumed to be periodic
with a fundamental ordering wave vector $\mathbf{G}$ plus its harmonics. The
period is assumed to be commensurate with the lattice, \textit{i.e.} there
exist integers $N$, $n$, and $m$ such that $N\mathbf{G=}\left( \frac{2\pi n}{%
a},\frac{2\pi m}{a}\right) $. %, where $N$, $n$, $m$ are integers).
(It is straightforward to extend the proof to the case, such as a
checkerboard CDW, in which there are two independent, commensurate ordering
vectors, $\mathbf{G}_1$ and $\mathbf{G}_2$, plus their combined harmonics.)
The system is described by the %mean-field
effective Hamiltonian %Bugoliubov-de Gennes Hamiltonian
\begin{equation}
\mathcal{H=H}_{0}+\mathcal{W}=\sum_{\mathbf{k}}\Psi_{\mathbf{k}}^{\dagger } %
\left[\mathrm{h}_{0}({\mathbf{k}})+\mathrm{w}{(\mathbf{k}})\right]\Psi_{%
\mathbf{k}}\ \ .  \label{Hfock}
\end{equation}
%Here $\mathcal{H}_{0}$ is the underlying translationally invariant
%Hamiltonian of a uniform d-wave superconductor and $\mathcal{W}$ is the
%perturbation. % with wavevector $\mathbf{G%
%}$. %$\mathbf{c}_{\mathbf{k}}$
For each ${\mathbf{k}}$ in the first Brillouin zone of the broken symmetry
state, $\mathrm{h}_{0}(\mathbf{k})$ and $\mathrm{w}({\mathbf{k}})$ are $2N
\times 2N$ matrices, and $\Psi_{\mathbf{k}}^\dagger$ is the $2N$ component
spinor operator,
\begin{equation}
\Psi_{\mathbf{k}}^\dagger = (\psi_{\mathbf{k}}^\dagger,\psi_{\mathbf{k}+%
\mathbf{G}}^\dagger,\psi_{\mathbf{k}+2\mathbf{G}}^\dagger, \ldots) \ ,
\end{equation}
%where
$\psi_{\mathbf{k}}^\dagger=\left( c_{\mathbf{k\uparrow }}^{\dagger },c_{-%
\mathbf{k\downarrow }}\right) $ is the usual Nambu spinor, and $c_{\mathbf{k}%
\sigma}^\dagger$ creates an electron in the Bloch state of the
non-interacting band with wave-vector $\mathbf{k}$ and spin polarization $%
\sigma$. %\begin{equation}
%\mathbf{c}_{\mathbf{k}}=\left(
%\begin{array}{c}
%\mathbf{c}_{\mathbf{k}} \\
%\mathbf{c}_{\mathbf{k}+\mathbf{G}} \\
%\mathbf{c}_{\mathbf{k}+2\mathbf{G}} \\
%\vdots%
%\end{array}%
%\right)
%\end{equation}
%
%where $\mathbf{c}_{\mathbf{k}}^{\dagger }=\left( c_{\mathbf{k\uparrow }%
%}^{\dagger },c_{-\mathbf{k\downarrow }}\right) $.

Since $\mathcal{H}_0$ is the effective Hamiltonian of a uniform nodal
superconductor, it is invariant under both time reversal, %transformation
$\mathcal{T}$ and translation, $\mathcal{S}_{\mathbf{R}}$, by any lattice
vector $\mathbf{R}$, and hence it is invariant under the combined symmetry
transformation, $\mathcal{T}_{\mathbf{R}} \equiv \mathcal{T}\mathcal{S}_{%
\mathbf{R}}$:
\begin{equation}
\mathcal{T}_{\mathbf{R}}^{-1}\mathcal{H}_0\mathcal{T}_{\mathbf{R}}=\mathcal{H%
}_0 .  \label{invar}
\end{equation}%
Taking into account the fact that $\mathcal{T}$ $\ $is anti-unitary, and so
satisfies $\mathcal{T}^{2}=-1$, it is straightforward to see that, with an
appropriate choice of basis,
\begin{equation}
\mathcal{T}_{\mathbf{R}}^{-1}\Psi _{\mathbf{k}}\mathcal{T}_{\mathbf{R}%
}=\Lambda _{\mathbf{R}}\Psi _{\mathbf{k}}^\star  \label{Sop}
\end{equation}%
where $\Lambda_{\mathbf{R}}$ is the tridiagonal unitary matrix
\begin{equation}
\Lambda _{\mathbf{R}}=\left(
\begin{array}{ccc}
i\mathbf{\sigma}_2e^{i\mathbf{k}\cdot \mathbf{R}} &  &  \\
& i \mathbf{\sigma}_2e^{i\left( \mathbf{k+G}\right) \cdot \mathbf{R}} &  \\
&  & \ddots%
\end{array}%
\right)  \label{lambda_mat}
\end{equation}%
and $\sigma _{2}$ is a Pauli matrix. Moreover, by the usual arguments, it
follows from the fermion anticommutation relations, Bloch's theorem, and
Kramer's theorem that $h_0(\mathbf{k})$ is a real, traceless, block diagonal
matrix. Specifically, % for $\mathbf{k}$
in the neighborhood of a nodal point, $\mathbf{k}_n$,
\begin{equation}
\mathrm{h}_{0}=\left(
\begin{array}{cc}
\mathrm{h}_{0}^{\prime } & 0 \\
0 & \mathrm{h}_{0}^{\prime \prime }%
\end{array}%
\right)
\end{equation}
where $\mathrm{h}_0^{\prime }({\mathbf{k}})$is a real, traceless, $2\times 2$
matrix which asymptotically has the Dirac form, $\mathrm{h}_0^{\prime }({%
\mathbf{k}}) \sim \left(\mathbf{k}-\mathbf{k}_n\right)\cdot \left ( \mathbf{v%
}_F\sigma_3 +\mathbf{v}_\Delta \sigma_1\right )$, and $\mathrm{h}%
_0^{\prime \prime }({\mathbf{k}})$ is a real, traceless $\left( 2N-2\right)
\times \left( 2N-2\right) $ matrix. ($\mathbf{v}_F$ and $\mathbf{v}_\Delta$
are, respectively, the Fermi velocity and the gap slope, and $\mathbf{v}%
_F\cdot \mathbf{v}_\Delta=0$.)

Thus far, we have simply reproduced the usual calculation of the
quasiparticle spectrum in an awkward basis with an artificially reduced
first Brillouin zone obtained by treating $\mathbf{G}$ as a reciprocal
lattice vector, and correspondingly we were forced to consider $N$ times as
many bands. The meaning of the statement that $\mathbf{G}$ does not nest the
nodal points is that, for $\mathbf{k}$ in the neighborhood of $\mathbf{k}_n$%
, all the eigenvalues of $\mathbf{h}_0^{\prime\prime}$ have a magnitude
larger than a non-zero, positive constant, $\Delta^{\prime\prime}$. The key
point is that $\mathrm{h}_0^\prime$ is a real, traceless matrix, so its
zeros have codimension 2, which means, generically, one can expect nodal
points in 2D or nodal lines in 3D.

We now proceed to demonstrate that if the perturbation, in addition to not
nesting the nodal points, is invariant under $\mathcal{T}_{\mathbf{R}}$ for
some $\mathbf{R}$, \textit{i.e.} $\mathcal{T}_{\mathbf{R}}^{-1} \mathcal{W}%
\mathcal{T}_{\mathbf{R}}=\mathcal{W}$, then the nodal point may move from $%
\mathbf{k}_n$ to a nearby point, $\tilde\mathbf{k}_n$, but it remains a
nodal point. From the invariance of $\mathcal{W}$ and Eq. (\ref{Sop}), it
follows that
\begin{equation}
\Lambda _{\mathbf{R}}^{\dagger }\mathrm{w}({\mathbf{k}}) \Lambda _{\mathbf{R}%
}=-\mathrm{w}({\mathbf{k}})  \label{Hk_cond}
\end{equation}
which immediately implies that $\mathrm{w}$ is traceless. To make further
progress, we express $\mathrm{w}$ and $\Lambda_{\mathbf{R}}$ in block form,
in the same way as we treated $\mathrm{h}_0$:
\begin{equation}
\mathrm{w}=\left(
\begin{array}{cc}
\mathrm{w}^{\prime } & \mathrm{u}^{\dagger } \\
\mathrm{u} & \mathrm{w}^{\prime \prime }%
\end{array}%
\right),~ \Lambda_{\mathbf{R}}=\left(
\begin{array}{cc}
\Lambda_{\mathbf{R}}^\prime & 0 \\
0 & \Lambda_{\mathbf{R}}^{\prime \prime }%
\end{array}%
\right)  \label{block}
\end{equation}%
where % $\mathrm{h}_0^{\prime }({\mathbf{k}})$ and
$w^{\prime }({\mathbf{k}})$ and $w^{\prime \prime }({\mathbf{k}})$ are,
respectively, a $2\times 2$
%matrices, $\mathrm{h}_0^{\prime \prime }({\mathbf{k}})$ and $%
%w^{\prime \prime }({\mathbf{k}})$ are
and a $\left( 2N-2\right) \times \left( 2N-2\right) $ Hermitian matrix, $u({%
\mathbf{k}})$ is a$\left( 2N-2\right) \times 2$ matrix, and $\Lambda_{%
\mathbf{R}}^\prime= i\sigma_2e^{i\mathbf{k}\cdot\mathbf{R}}$.
%From Eq. \ref{block} it follows that $\{\sigma_2,\mathrm{w}^\prime\}=0$, which in turn implies $\mathrm{w}^\prime$ is real.
%
%We order the components so that $\mathrm{h}^{\prime }\left( {\mathbf{k}}\right)
%$ vanishes at the nodal point $\mathbf{k}_{n}$ of $\mathrm{h}_{0}$, \textit{%
%i.e.} $\mathrm{h}^{\prime }({\mathbf{k}})\sim ({\mathbf{k}}-{\mathbf{k}}%
%^{\prime })\cdot \mathbf{v}$ where $\mathbf{v}$ is a $2\times 2$ matrix
%valued vector. Formally, the statement that the secondary order does not "nest" the nodal points is equivalent to the statement that the eigenvalue spectrum of $\mathrm{h}_0^{\prime\prime}(\mathbf{k}_n) is gapped.  Consequently,
%Finally, s
Since $\mathrm{h}_0^{\prime\prime}(\mathbf{k}_n)$ is gapped, for a small
enough perturbation, the matrix $\mathrm{h}_{0}^{\prime \prime }+\mathrm{w}%
^{\prime \prime }$ has an inverse and hence the low energy states in the
region of ${\mathbf{k}}$-space near ${\mathbf{k}_n}$ can be found \textit{%
asymptotically exactly} (\textit{i.e.} neglecting errors that vanish in
proportion to $E/\Delta^{\prime\prime}$) by diagonalizing the effective $%
2\times 2$ Hamiltonian
\begin{equation}
\mathrm{h}_{\mathrm{eff}}^\prime\left( \mathbf{k}\right) \equiv\mathrm{h}%
_{0}^{\prime }+\mathrm{w}^{\prime }-\mathrm{u}^{\dagger }\left( \mathrm{h}%
_{0}^{\prime \prime }+\mathrm{w}^{\prime \prime }\right) ^{-1}\mathrm{u}.
\label{heff}
\end{equation}%
In particular, at a nodal point of the full Hamiltonian, $\mathrm{h}_{%
\mathrm{eff}}(\mathbf{k})$ must have a zero eigenvalue. It is
straightforward to see from Eq. (\ref{Hk_cond}) that
\begin{equation}
[\Lambda _{\mathbf{R}}^\prime]^{\dagger }\mathrm{h}_{\mathrm{eff}} \Lambda _{%
\mathbf{R}}^\prime=\sigma_2\mathrm{h}_{\mathrm{eff}}\sigma_2 = -\mathrm{h}_{%
\mathrm{eff}}.  \label{hefflambda}
\end{equation}
This, combined with the condition that it be Hermitian, implies that $%
\mathrm{h}_{\mathrm{eff}}$ is also a real, traceless matrix, $\mathrm{h}_{%
\mathrm{eff}}\left( \mathbf{k}\right) =a\left( \mathbf{k}\right) \sigma
_{1}+b\left( \mathbf{k}\right) \sigma _{3}$ where $a\left( \mathbf{k}\right)
$ and $b\left( \mathbf{k}\right) $ are real functions.

The robustness of the node follows directly. %Eq. (\ref{heff}) is
%satisfied when
The quasiparticle spectrum derived from $\mathrm{h}_{\mathrm{eff}}(\mathbf{k}%
)$ is $E(\mathbf{k}) = \pm \sqrt{a^2(\mathbf{k})+b^2(\mathbf{k})}$, which
vanishes only when $a\left( \mathbf{k}\right) = b\left( \mathbf{k}\right) =0$%
. Since we have two tuning parameters, $k_{x}$ and $k_{y}$, a solution
generically exists. More specifically, for $\mathbf{k}$ near $\mathbf{k}_n$,
$a(\mathbf{k}) = \mathbf{v}_\Delta\cdot (\mathbf{k}-\mathbf{k}_n) + \delta a$
and $b(\mathbf{k}) = \mathbf{v}_F\cdot (\mathbf{k}-\mathbf{k}_n) + \delta b$
where $\delta a$ and $\delta b$ are small so long as the perturbation is
weak; the node simply moves a small distance in $\mathbf{k}$-space.
%starting from a nodal point of $\mathrm{h}%
%_{0}\left( \mathbf{k}\right) $ and adding a small perturbation $\mathrm{w}%
%\left( \mathbf{k}\right) $, the two contours of $a\left( \mathbf{k}\right) =0
%$ and $b\left( \mathbf{k}\right) =0$ are deformed continuously, so their
%crossing points can shift, but they can not generically disappear unless
%they annihilates with another crossing point. Therefore t
The nodes must be stable for a finite range of strength of the perturbation,
which concludes our proof. Conversely, for a perturbation that nests the
nodal points, or for which no symmetry of the form of $\mathcal{T}_{\mathbf{R%
}}$ exits, our proof breaks down, which suggests, but does not prove, that
the nodal structure is fundamentally altered by this sort of symmetry
breaking. This argument is related to the Wigner-von Neumann theorem\cite%
{WignerVN}.
%that states that generically, in order to get a level crossing point of a
%general real, symmetric matrix, such as $\mathrm{h}_{\mathrm{eff}}\left(
%\mathbf{k}\right) $, two independent parameters need to be tuned.

%\begin{equation}
%H_{0,\mathbf{k}}=\left(
%\begin{array}{cc}
%\xi _{\mathbf{k}} & \Delta _{\mathbf{k}} \\
%\Delta _{\mathbf{k}} & -\xi _{\mathbf{k}}%
%\end{array}%
%\right)
%\end{equation}

%Next, let us illustrate the above principle by solving the Bugoliubov-de
%Gennes (BdG)
\textit{Examples} - To illustrate the above principle, we explicitly compute
the QP spectrum of the mean-field Hamiltonian of a d-wave superconductor in
the presence of various symmetry-breaking orders that have been considered
in the context of the cuprate high temperature superconductors. %The
%unperturbed Hamiltonian is taken to be
%in the simple d-wave BdG form, with
To be explicit, we assume an underlying band-structure such that $\xi _{%
\mathbf{k}}=-2t\left( \cos k_{x}+\cos k_{y}\right) -4t^{\prime }\cos
k_{x}\cos k_{y}-\mu $. The parameters used are $t=1$, $t^{\prime }=-0.25$
and $\mu =-0.9$, representing a generic cuprate-like band structure. The gap
function is taken to be
\begin{equation}
\Delta _{\mathbf{k}}=\Delta _{0}\left( \cos k_{x}-\lambda \cos k_{y}\right)
\label{gapeq}
\end{equation}%
with $\Delta _{0}=0.4$ and $\lambda =1$. (In some cases, we will explore the
effect of an orthorhombic distortion, which we will incorporate by letting $%
\lambda \neq 1$.)

\begin{figure}[t]
\centering
\includegraphics[width=7cm]{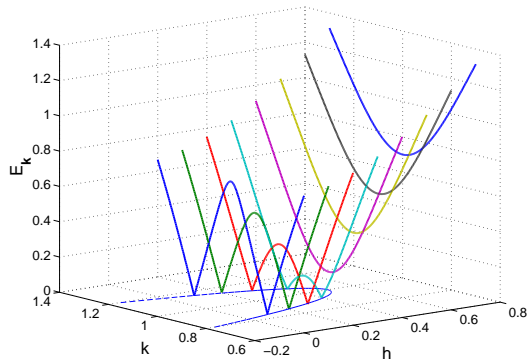}
\caption{(color online.) Spectrum of a d-wave superconductor along the line $%
(k_x,k_y)=(\frac{\protect\pi}{2},\frac{\protect\pi}{2})k$ in the presence
coexisting SDW order (\textit{i.e.} a staggered Zeeman field) of magnitude $%
h $. The nodal points are stable until $h \approx 0.45$, when two nodal
points meet in k-space and a gap opens.}
\label{fig:sdw}
\end{figure}

As a first example, let us consider a $\mathbf{G}=\left( \pi ,\pi \right) $
spin density wave, represented by the perturbation $\mathcal{W}_{\text{SDW}%
}=h\sum_{\mathbf{k},\sigma }\sigma c_{\mathbf{k}+\mathbf{G},\sigma
}^{\dagger }c_{\mathbf{k},\sigma }$.
% where $\sigma =\pm 1$ is the spin index.
This perturbation manifestly breaks time reversal symmetry. However, it is
invariant under time reversal %symmetry
followed by a translation by $\mathbf{R}=a\mathbf{\hat{x}}$. Diagonalizing
the % BdG
effective Hamiltonian numerically we find that the nodes in this case are
robust. Upon increasing $h$, the nodes are shifted from their original
position. Due to the reflection symmetry of the perturbation around the $%
\pi\left( 1 ,1 \right) $ direction, the nodes are constrained to move along
the $\left( 0,0\right) $ to $\pi\left( 1 ,\pm 1 \right) $ lines. When $%
h\approx 0.45$, the nodes reach the points $\pm \frac{\pi }{2} \left(
1,1\right) $, and so are nested by the ordering vector $\mathbf{G}$.
% nests node at $\left( -\frac{\pi }{2},-\frac{\pi }{2}\right) $ and a gap opens.
For $h > 0.45$, the spectrum is fully gapped. The spectrum for various
values of $h$ is shown in Fig. \ref{fig:sdw}
%shows the spectrum along the line $%\left(
%k_{x},k_{y}\right)
along the line $\mathbf{k}= \frac{\pi }{2}\left(k,k\right) $.
%various values of $h$.

\begin{figure}[b]
\centering
\includegraphics[width=6cm]{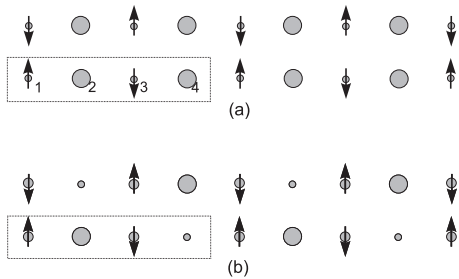}
\caption{Two patterns of unidirectional spin and charge order (``stripes'')
discussed in the text. The arrows represent the spin density, and the size
of the circles represents the charge density. The rectangles are the unit
cells. The primitive vectors are $(4a,0)$ and $(2a,a)$.}
\label{fig:stripes}
\end{figure}

\begin{figure}[t]
\centering
\includegraphics[width=6cm]{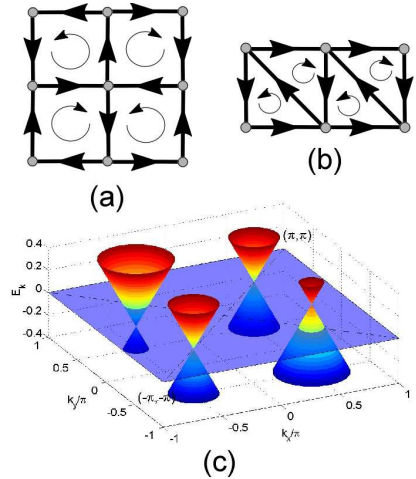}
\caption{(color online.) (a,b) Two patterns of spontaneous orbital currents
on a square lattice. (a) the $(\protect\pi,\protect\pi)$ d-density wave
pattern, and (b) is a square lattice version of the $(0,0)$ order proposed
in \onlinecite{Varma}. (c) Qualitative low energy spectrum of a d-wave
superconductor with coexisting orbital current order of type (b). Note that
the pair of nodes along the line $k_x=k_y$ remain at zero energy, while the
nodes along $k_x=-k_y$ are shifted away from zero energy in opposite
directions, forming hole-like or electron-like pockets.}
\label{fig:cones}
\end{figure}

To illustrate further the special role played by translation symmetry, we
will next consider the two combinations of %unidirectional
spin and charge density waves (\textquotedblleft stripes\textquotedblright
), shown in Fig. \ref{fig:stripes} a \& b. In both cases, time-reversal and
translation symmetry are broken, such that there are 4 sites in the new unit
cell. However, in state-a, time reversal followed by translation by $\mathbf{%
R}=2a\hat\mathbf{x}$, or by $\mathbf{R}=a\hat\mathbf{y}$, %4 sites in the
%$x$ direction, \textit{i.e.} $\mathcal{T}_{4\hat\mathbf{x}}$,
remains an unbroken symmetry, whereas in state-b, no symmetry of the form of
$\mathcal{T}_{\mathbf{R}}$ survives. (There do, however, remain unbroken
symmetries which combine $\mathcal{T}$ and reflections through a plane.) We
represent these states by % patterns correspond to the
a perturbation Hamiltonian of the form: $\mathcal{W}%_{a,b}
=\sum_{\mathbf{r}\alpha }\left( V_{\alpha }%^{(a,b)}
n_{\mathbf{r},\alpha }-h_{\alpha }%^{(a,b)}
S_{\mathbf{r},\alpha }^{z}\right) $, where $\mathbf{r}$ is the
%unit cell index,
Bravais lattice vector labelling a unit cell, $\alpha =1,..,4$ is the index
of the basis site in each unit cell, and $n_{\mathbf{r},\alpha
}=\sum_{\sigma }c_{\mathbf{r},\alpha ,\sigma }^{\dagger }c_{\mathbf{r}%
,\alpha ,\sigma }$ and $S_{\mathbf{r},\alpha }^{z}=\frac{1}{2}\sum_{\sigma
}\sigma c_{\mathbf{r},\alpha ,\sigma }^{\dagger }c_{\mathbf{r},\alpha
,\sigma }$ are the local charge and spin densities, respectively. Following
the site labelling scheme shown in Fig. \ref{fig:stripes}, %(a),
we take represent the effective field conjugate to the spin density by the 8
component vector $h%_{\alpha
%=1,...,8}%^{(a)}
=\left( h,0,-h,0\right) $. In case a, the field conjugate to the density is
%we have in case a)
$V%_{\alpha
%=1,...,8}%^{(a)}
=\left( 0,-V,0,-V\right) $ %and $h_{\alpha
%=1,...,8}%^{(a)}
%=\left( h,0,-h,0,-h,0,h,0\right) $ for pattern (a),
while for case b, $V%_{\alpha =1,...,8}^{(b)}
=\left( 0,-V,0,V\right) $. % and
%$h_{\alpha }^{(b)}=h_{\alpha }^{(a)}$ for pattern (b). Note that
%pattern (a) is invariant
%under time reversal followed by a translation by $\mathbf{R}=4a\mathbf{\hat{x%
%}}$, while pattern (b) does not have any such symmetry property.
%((b) is invariant under time reversal followed by a
%\textit{reflection}, but that does not protect the nodes.)
%Therefore, We expect that the nodes will be robust for a finite
%range of parameters in (a), while in (b) they will be removed by
%any infinitesimal stripe perturbation. Indeed, from the numerical
%diagonalization the BdG Hamiltonian, we find that the nodes in (a)
We have computed the spectrum numerically for fixed $h$ as a function of
increasing $V$. As expected on the basis of our general theorem, in case-a
the nodal points survive until $V$ exceeds a critical value of order unity.
Conversely, in case-b, where no general theorem insures the stability of the
nodes, we find that
%remain up to values of $h$ and $V$ of order unity, while for (b)
%they are removed immediately and
a gap opens for arbitrarily small $V$ and grows as %. For fixed $V$, the
%gap behaves as
$\Delta \sim V^{2} $.

The last example we will consider is the case of spontaneous orbital current
loops. Fig. \ref{fig:cones}(a,b) shows two orbital current patterns: the
first (a) is known as \textquotedblleft d-density wave\textquotedblright\
(dDW)\cite{DDW} for which $\mathbf{G}=\left( \pi ,\pi \right) $ , and the
second (b) is a $\mathbf{G}=\left( 0,0\right) $ pattern which, from a broken
symmetry viewpoint, is equivalent to a state defined by Varma on the
somewhat more complex Cu-O lattice % is a square
%lattice version of the pattern proposed in [\onlinecite{Varma}] for the CuO$%
%_{2}$ plane (in the sense that it has the same symmetries as those of the
%pattern
in [\onlinecite{Varma}]. For either of these states, the perturbation
Hamiltonian is % written as
of the form $\mathcal{W}%_{a,b}
=-i\sum_{\mathbf{rr}^{\prime }\sigma }J_{\mathbf{rr}^{\prime }}%^{a,b}
c_{\mathbf{r}\sigma }^{\dagger }c_{\mathbf{r}^{\prime }\sigma }+H.c.$, where
the connectivity of the current network $J_{\mathbf{rr}^{\prime }}%^{a,b}
$ is determined according to Fig. \ref{fig:cones}(a,b), and all the non-zero
currents have the same magnitude $J$. Based on our general principle, we
expect that in pattern (a) the superconducting nodes will survive at least
over a finite range of $J$, since it is invariant under time reversal
followed by a translation by $\mathbf{R}=a\mathbf{\hat{x}}$. Pattern (b), on
the other hand, is a $\mathbf{G}=\left( 0,0\right) $ pattern, so the\ nodes
may be removed immediately.

In fact, for pattern (a), $W_{\mathbf{k}}=0$ at the nodal points (since it
has d-wave symmetry), so the nodes are trivially stable. To avoid this
non-generic situation, we introduce orthorhombicity of the gap function by
setting $\lambda =0.5$ in Eq. (\ref{gapeq}). Diagonalizing the BdG
Hamiltonian numerically for this case, we still find that the nodes exist up
to values of $J$ of the order of the bandwidth $t$. Since pattern (b) does
not break translation symmetry, %so
the corresponding %BdG
effective Hamiltonian can be easily diagonalized. The eigenenergies are $%
E_{\pm ,\mathbf{k}}^{\left( b\right) }=\pm \sqrt{\xi _{\mathbf{k}%
}^{2}+\Delta _{\mathbf{k}}^{2}}+2J\left\{ \sin \left( k_{x}a\right) -\sin
\left( k_{y}a\right) +\sin \left[ \left( k_{y}-k_{x}\right) a\right]
\right\} $. We see that even though the degeneracy at the nodes is not
lifted, some of the nodes (the ones that lie on the line $k_{x}=-k_{y}$) are
shifted away from zero energy: if $J>0$, then the node at $k_{x}>0$ ($%
k_{x}<0 $) is shifted to positive (negative) energy, respectively. The
low-energy spectrum is shown qualitatively in Fig. \ref{fig:cones}(c). In
the case of an orthorhombic gap function ($\lambda \neq 1$), the nodes along
$k_{x}=k_{y} $ are also shifted to finite energies by an amount proportional
to $\lambda -1$. Whenever a node is shifted to a finite energy, a hole-like
or electron-like pocket is formed, giving rise to a finite density of states
at zero energy.

\emph{Discussion -} The condition established here for the perturbative
stability of the nodal points is a sufficient condition, not a necessary
one. However, from the examples considered above, we see that in several
cases where the condition is not met, the nodes are %removed (
gapped or shifted to finite energies upon introducing an infinitesimal
perturbation. Therefore, we expect that in generic cases, this condition is
%essentially
more or less necessary for the existence of nodal points. Thus, we can
classify various weakly ordered states according to whether they leave the
QP nodes intact (\textit{e.g.} dDW or certain types of stripes), gaps them (%
\textit{e.g.} density wave order with ordering vector $\mathbf{G}$ which
nests the nodal points), or moves them away from the Fermi energy (\textit{%
e.g.} Varma loops). Since the existence of nodal points has several
well-defined experimental consequences, this information can %principle can
%help to detect indirectly
provide a non-trivial consistency check on the assumed occurrence of various
forms of coexisting order in the superconducting phase.
%, by looking for the removal
%of the nodes. Conversely,
In particular, in situations where the experiments are consistent with nodal
QPs, one can rule out (or at least give an upper bound on) competing orders
which are not invariant under time reversal followed by translation. For
instance, measurements of the linear in $T$ decay of the superfluid density
in ultra-pure crystals of the Ortho-II phase of YBCO\cite{lin_T_lambda} give
evidence that the nodal points are gapless and tied to the Fermi energy to
within an accuracy of approximately 1K, or in other words a fraction of a
percent of $\Delta_0$. This can probably be converted into a rather
stringent bound on the strength of Varma loop order at low temperatures in
this material.

%These observations are particularly interesting in the context of the recent
%Kerr effect measurements\cite{AharonKerr} in YBCO-123, which detect a finite
%signal with an onset temperature close to the pseudogap temperature. The
%signal extends well into the superconducting phase. These experiments, which
%are consistent with earlier $\mu$SR\cite{mSR} and polarized neutron
%scattering\cite{Bourges} measurements, imply that a time reversal symmetry
%breaking order exists in YBCO, and that this new type of order coexists with
%the superconducting order at least in some region of the phase diagram.
%As
%discussed above, the stability of the nodes can be used to give theoretical
%constraints on the type or magnitude of this order.

The extension of these results to the case of an incommensurate perturbation
is an interesting open problem.
%, to the best of our knowledge, still unclear. %This
%is an important issue, since there is at least some evidence for
%incommensurate charge and spin order \cite{incomm} in the cuprate
%phase diagram.

A ``striped superconducting'' state\cite{us}, which can be thought of as a
spontaneously developed FFLO state, was considered as a candidate state to
explain some extremely anomalous transport data\cite{tranquada} that was
recently obtained on the stripe-ordered material La$_{2-x}$Ba$_x$CuO$_4$. In
the proposed state, the superconducting order parameter is modulated with
wavevector $\mathbf{G}=(\pi/4,0)$. A calculation of the spectrum in this
state reveals that the nodes expand to a series of pockets along the bare
Fermi surface, and the single particle spectral function has low energy
weight along finite Fermi arcs. The reason that the nodes are not protected,
%in this state, despite the
%presence of
despite the time reversal symmetry of this state, %is that this state
is that it cannot be %considered to be
viewed as a %slightly
perturbed version of a uniform d-wave state.
%(\textit{i.e.}, the perturbation is ``large"). %The resulting Fermi
%arcs and the finite density of states at zero energy can be used
%to detect this state.

\textit{Acknowledgments:} We thank T.L. Hughes, S. Raghu, E. W.
Carlson and O. Vafek for useful comments. One of us (CCC) is
supported in part by National Science Council (NSC), Taiwan, under
grant NSC-095-SAF-I-564-013-TMS. This work was supported by NSF
grant \# DMR-0531196 at Stanford.

\end{document}